\documentclass[preprint]{elsarticle}

\usepackage{graphicx}
\usepackage{lineno,hyperref}
\usepackage{amsmath,amsthm,amssymb}
\usepackage{url}
\usepackage{booktabs}
\usepackage{multirow}

\begin{document}

\begin{frontmatter}

\title{Community detection based on first passage probabilities}

\author[1,2]{Zhaole Wu}
\author[1,2,3,5]{Xin Wang}
\author[1,4,5]{Wenyi Fang}
\author[1,2,5,6]{Longzhao Liu}
\author[1,2,5]{Shaoting Tang\corref{mycorrespondingauthor1}}
\cortext[mycorrespondingauthor1]{Corresponding author}
\ead{tangshaoting@buaa.edu.cn}
\author[7]{Hongwei Zheng\corref{mycorrespondingauthor1}}
\ead{hwzheng@pku.edu.cn}
\author[1,2,5,7]{Zhiming Zheng}

\address[1]{LMIB, NLSDE, BDBC, Beijing, 100191, China}
\address[2]{School of Mathematical Sciences, Beihang University, Beijing 100191, China}
\address[3]{Department of Mathematics, Dartmouth College, Hanover, NH 03755, USA}
\address[4]{School of Mathematical Sciences, Peking University, Beijing, 100871, China}
\address[5]{PengCheng Laboratory, Shenzhen, 518055 , China}
\address[6]{ShenYuan Honor School, Beihang University, Beijing 100191, China}
\address[7]{Institute of Artificial Intelligence and Blockchain, Guangzhou university, Guangdong province 510006, China}

\begin{abstract}
Community detection is of fundamental significance for understanding the topology characters and the spreading dynamics on complex networks. While random walk is widely used and is proven effective in many community detection algorithms, there still exists two major defects: (i) the maximal length of random walk is too large to distinguish the clustering information if using the average step of all possible random walks; (ii) the useful community information at all other step lengths are missed if using a pre-assigned maximal length. In this paper, we propose a novel community detection method based on the first passage probabilities (FPPM), equipped with a new similarity measure that incorporates the complete structural information within the maximal step length. Here the diameter of the network is chosen as an appropriate boundary of random walks which is adaptive to different networks. Then we use the hierarchical clustering to group the vertices into communities and further select the best division through the corresponding modularity values. Finally, a post-processing strategy is designed to integrate the unreasonable small communities, which significantly improves the accuracy of community division. Surprisingly, the numerical simulations show that FPPM performs best compared to several classic algorithms on both synthetic benchmarks and real-world networks, which reveals the universality and effectiveness of our method.

\end{abstract}


\end{frontmatter}


\section{Introduction}

The structural characters greatly shape the expression of functions on various complex systems, ranging from metabolic networks, brain networks to socio-economic networks ~\cite{barabasi2016network,jonsson2006cluster,jalan2017unveiling,alexander2012discovery,wu2011overlapping,zachary1977information,boccaletti2006complex,girvan2002community}. With the rapid development of large-scale social networks, the clustering structures (i.e., communities) have shown great importance in many spreading processes, such as the formation of social polarization~\cite{wang2020public}, the diffusion of competitive information~\cite{liu2020homogeneity} and the identification of influential global spreaders~\cite{hu2018local}. As all the understandings of these dynamical evolutions are based on the quality of community partition, community detection has become increasingly important and has attracted great attention recently~\cite{ullah2017community,liu2018global,runge2019detecting,zhu2019community,chen2018novel,huang2019community,boccaletti2007detecting}. Generally speaking, communities are groups of vertices where edges that represent the existence of relationship between vertices within each community are more concentrated than those bridging different communities.

Initially, the community detection methods were mostly proposed in view of optimization~\cite{newman2004fast,newman2004finding,clauset2004finding}, spectrum~\cite{white2005spectral,capocci2005detecting,newman2013spectral,krzakala2013spectral} and statistic~\cite{hastings2006community,newman2007mixture,copic2009identifying}, while few attention has been paid on the relevance between community structure and spreading dynamics. Subsequently, random walk methods are proposed based on the following facts: on one hand,  the community structure results in a longer time for random walk to spend inside the communities; on other hand, the trace of random walk contains the crucial information about community structures~\cite{feller1957introduction,masuda2017random,fortunato2010community,fortunato2016community}. Zhou characterized each vertex with a vector where the elements are the average numbers of edges that a random walker has to cross from the vertex to all the other vertices, and defined the vertex distances as the euclidean distances between the vectors~\cite{zhou2003distance}. Then a divisive procedure is used to divide networks into communities by a certain threshold. In another work, Zhou and Lipowsky proposed a new method called Netwalk~\cite{zhou2004network}. They defined a proximity index between vertices based on the mean first passage time~\cite{paul2013stochastic}. Once the distances or the similarities between vertices are set up, the problem of community detection is simplified to a problem of clustering, where a great number of effective methods have been proposed such as k-means~\cite{hartigan1979algorithm}, density-based clustering~\cite{ester1996density} and hierarchical clustering~\cite{mullner2011modern}. These works provide effective community divisions on some real networks and artificial benchmarks. However, the above methods use the average number of steps that consider all length of possible random walks, which is often too large to distinguish the corresponding topology information, as the random walk will converge to a stationary state when the number of steps approaches infinity.

Another series of methods utilizing random walk pre-assign a small fixed number of steps that the random walk can move. A well-known one was introduced by Latapy and Pons~\cite{pons2005computing}. They proposed a new distance between vertices: two vertices are close if the probabilities that a random walker moves from the two vertices to others at a fixed number of steps are similar. Besides, Hu et al. considered the signal propagation where the signaling process from each vertex with a fixed length of time is used to measure the influence of this vertex among the whole network, and vertices with the same community are expected to have similar influences~\cite{hu2008community}. The main weakness for these methods lies on the fact that there is no universal length of random walk for all the networks. What's more, they only consider the statistical properties of random walk at the pre-assigned length while possibly missing useful information at other step lengths.

In this paper, we propose a novel community detection method based on the first passage probabilities named as FPPM~\cite{paul2013stochastic}, which provides an effective solution for the defects mentioned above. Instead of using the average step length or a pre-assigned step length, our algorithm chooses diameter of the network as an appropriate maximal length of random walks which is adaptive to different networks. Furthermore, we design a new similarity measure which calculates the sum of weighted correlations between first passage probabilities of random walk at multiple times and incorporates more topology information. Then the hierarchical clustering is used to group vertices into communities, within which a best division is chosen via modularity value. Finally, we conduct a post-processing strategy inspired by LPA (label propagation algorithm) to eliminate the unreasonable small communities, which improves the precision of the final division. Results show that on both artificial and real-world networks, FPPM outperforms several classic and well-known algorithms, including Fastgreedy, Infomap, LPA, Louvein and Walktrap.

\section{Preliminaries}
Given a connected network $G = (V, E)$ where $V=\{v_1, v_2, ..., v_N\}$ is a set of $N$ vertices and $E$ is the set of $M$ edges. The adjacent matrix, A, is an $N \times N$ matrix where $A_{ij}=1$ if $v_i$ and $v_j$ are connected and $A_{ij}=0$ otherwise. For simplicity and without loss of generality, we only use unweighted and undirected networks in this paper. The vertex set $\mathcal{N}_i=\{v_j|A_{ij}\neq 0\}$ is called the neighbors of $v_i$. The size of $\mathcal{N}_i$ is defined as the degree of $v_i$ and denoted as $k_i$. The community structure of a network $G$ is represented as a partition of $V$, namely
\begin{equation}
    \mathcal{C}=\{C_1, C_2, ..., C_k| C_i \cap C_j = \emptyset \wedge \bigcup_{i=1}^{k}C_i=V\}
\label{1}
\end{equation}
where the community partition $\mathcal{C}$ is a set of communities.

Consider a discrete random walk on a connected network $G$~\cite{masuda2017random}. A walker randomly moves from the current vertex to one of its neighbors at each time step. The transition matrix is denoted as $T$, and $T_{ij}$ is the probability that the walker moves from $v_i$ to $v_j$. Clearly, $T$ satisfies:
\begin{equation}
    \begin{aligned}
    \sum_{j=1}^N T_{ij}=1, &\forall \; i \\
    T_{ij} \geq 0, &\forall \; i,j
    \end{aligned}
\label{2}
\end{equation}

Define the current vertex where the walker stops at time $t$ as $s_t$. The conditional probabilities $f_{ij}^{(n)}=P(s_n=v_j \wedge s_t\neq v_j, \forall \; 0<t<n|s_0=v_i)$ are the n-step first passage probabilities ($i\neq j$) or the n-step first return probabilities ($i=j$) from $v_i$ to $v_j$~\cite{paul2013stochastic}. For convenience, we use n-step first passage probabilities to refer to both cases. According to the conditional probability equation~\cite{feller1957introduction}, $f_{ij}^{(n)}$ can be written as:
\begin{equation}
    f_{ij}^{(n)} = \sum_{k=1, k\neq j}^{N} T_{ik} \cdot f_{kj}^{(n-1)}
\label{4}
\end{equation}

In the matrix notation, we denote the n-step first passage probabilities matrix as $F^{(n)}$ where $F_{ij}^{(n)}=f_{ij}^{(n)}$. Obviously, $F^{(1)}$ is equal to the transition matrix $T$. Combined with Eq. (\ref{4}), the first passage probabilities meet the following iterative formulas:
\begin{equation}
    \begin{aligned}
    F^{(n+1)} &= T \cdot (F^{(n)} - F_{dg}^{(n)})\\
    F^{(1)} &= T
    \end{aligned}
\label{5}
\end{equation}
where $F_{dg}^{(n)}$ is the diagonal matrix of $F^{(n)}$.

\section{Algorithm}
In this section, we first introduce the detailed three components of FPPM: 1) a specific discrete random walk with a novel measure of vertex similarity based on the first passage probabilities, 2) a hierarchical agglomerative clustering algorithm to get a dendrogram, 3) a subsequent processing inspired by LPA to get the final community partition. Finally, we calculate the time complexity of our algorithm.

\subsection{Discrete random walk}
Intuitively, the edges inside the same communities are more than those joining different communities. Therefore, edges between communities are included in very few or even no triangles ~\cite{radicchi2004defining}. In other words, two adjacent vertices in the same community are more likely to have common neighbors compared to adjacent vertices in different communities. Conversely, the more common neighbors a pair of adjacent vertices share, the more probable they are in the same community. Based on these insights, we design a specific random walk where the transition probability is proportional to the number of common neighbors the two vertices have. The transition matrix is as follows:
\begin{equation}
    T_{ij}=\left\{
    \begin{aligned}
    &\frac{| \mathcal{N}_i \cap \mathcal{N}_j | + 1}{\sum_{j\in \mathcal{N}_i} (| \mathcal{N}_i \cap \mathcal{N}_j | + 1)} &, j \in \mathcal{N}_i \\
    &0 &, j \notin \mathcal{N}_i
    \end{aligned}
    \right.
\label{6}
\end{equation}
here we add one in the numerator to prevent the special case where a vertex has no common neighbors with all its neighbors. Then we can calculate the corresponding $F^{(n)}$ by Eq. (\ref{5}) and characterize every vertex $v_i$ with a vector $F_{i\cdot}^{(n)}$, the $i^{th}$ row of $F^{(n)}$ at time $n$ . As the vertices within the same community exhibit similar characteristics revealed by random walk, the n-time similarities between pairs of vertices can be defined as follows:
\begin{equation}
    s_{ij}^{(n)} = \frac{\sum_{k=1}^{N} (F_{ik}^{(n)} - \bar{F}_{i\cdot}^{(n)})(F_{jk}^{(n)} - \bar{F}_{j\cdot}^{(n)})}{\sigma(F_{i\cdot}^{(n)}) \cdot \sigma(F_{j\cdot}^{(n)})}
\label{7}
\end{equation}
where
\begin{equation}
    \begin{aligned}
        \bar{F}_{i\cdot}^{(n)} &= \frac{1}{N}\sum_{k=1}^{N}F_{ik}^{(n)} \\
        \sigma(F_{i\cdot}^{(n)}) &= \sqrt{\sum_{k=1}^{N}(F_{ik}^{(n)} - \bar{F}_{i\cdot}^{(n)})^2}
    \end{aligned}
\label{8}
\end{equation}

In the end, the similarities are defined as the weighted sum of n-time similarities:
\begin{equation}
    s_{ij} = \frac{\sum_{n=1}^{n_{max}} w_n\cdot s_{ij}^{(n)}}{\sum_{n=1}^{n_{max}}w_n}
\label{9}
\end{equation}
where $w_n$ is the weight coefficient and $n_{max}$ is the maximal step that the walker can take. We collect the similarities between all pairs of vertices into an $N\times N$ matrix denoted as $S_v$. In numerical experiments, we take linear weight coefficient where $w_n=n-1$ which prefers longer walks as they collect more structural information. Besides, we let $w_1$ be zero and exclude $s_{ij}^{(0)}$ since $F^{(1)}=T$ is a sparse matrix which may lead to low similarities between vertices within the same communities.

In addition, the maximal step is decided based on the following reasons. On one hand, a walker setting out from every vertex can reach all other vertices sooner or later in a connected networks, i.e., $\sum_{n=1}^{\infty} F_{ij}^{(n)} = 1,\forall \;i,j$, which leads to $\lim_{n\to \infty} F_{ij}^{(n)} = 0, \forall \;i,j$. As a consequence, two vertices belonging to different communities may have a high similarity if $n$ is too large. On the other hand, the random walk is not able to effectively reflect the information about network topology if $n$ is too small. Hence, here we choose the diameter of network $G$ as an appropriate maximal step $n_{max}$, which is simple and adaptive of all networks. Our method guarantees the walker to reach every vertex and capture the complete community information. Meanwhile, the trace of the walker would not be too long. The numerical simulations suggest that our choice works well.

\subsection{Hierarchical agglomerative clustering}
With the vertex similarities derived, the problem of community detection is transformed into a problem of clustering. Our FPPM uses hierarchical agglomerative clustering (HAC) to group vertices into communities~\cite{mullner2011modern}. HAC is a bottom-up clustering algorithm, which initializes the community partition $\mathcal{C}$ as $\{\{v_1\},\{v_2\},...,\{v_N\}\}$. As the algorithm goes on, two "closest" communities will be merged into a new community continuously until only one community left. The key of HAC is how to quantify the "closeness"  between communities. In the previous section, we have introduced a similarity function between pairs of vertices. Now we extend it to a measure of community similarities. Supposing $C_i$ and $C_j$ to be two distinct communities in $\mathcal{C}$, we define the similarity between $C_i$ and $C_j$ as the average of similarities between vertices in $C_i$ and vertices in $C_j$:
\begin{equation}
    S_{C_i,C_j}=\frac{1}{|C_i|\cdot |C_j|}\sum_{v_m\in C_i,v_n\in C_j}s_{mn}
\label{10}
\end{equation}
where $S_{C_i,C_j}=s_{ij}$ for the initial partition $\{\{v_1\},\{v_2\},...,\{v_N\}\}$. With this similarity definition, we set off from the initial partition and initialize the community similarly matrix $S$ where $S_{ij}=S_{C_i,C_j}$. Afterwards, our method performs the following operations repeatedly until there is only one community in the partition $\mathcal{C}$:
\begin{itemize}
    \item Choose the pair of communities $C_i$ and $C_j$ with the maximal similarity from $\mathcal{C}$.
    \item Remove $C_i$ and $C_j$ from $\mathcal{C}$, merge them into a new community $C_k = C_i \cup C_j$ and append $C_k$ to $\mathcal{C}$.
    \item Update the similarities between $C_k$ and other communities remained in $\mathcal{C}$.
\end{itemize}
In the last procedure, the new similarity between $C_k$ and another community $C_l$ can be easily calculated by the following updating formula:
\begin{equation}
    S_{C_k, C_l}=\frac{|C_i|}{|C_i|+|C_j|}S_{C_i,C_l}+\frac{|C_j|}{|C_i|+|C_j|}S_{C_j,C_l}
\label{11}
\end{equation}

When HAC procedure ends, we obtain a dendrogram (binary tree) composed of $N$ partitions. The root of the dendrogram is the partition that consists of only one community containing all vertices, and all the leaves make up the initial partition. The internal node in the dendrogram represents the community which is produced by merging its two children nodes. The dendrogram encompasses the information of all hierarchical community structure, from which we can get an ideal partition of community structure.

\subsection{Post-processing}

From top to bottom, the dendrogram is composed of a series of nested community partitions from coarse to fine, and the "coarsest" and "finest" ones are two trivial partitions. How to choose the "best" partition from the dendrogram? Modularity is a widely used quality function of the strength of community structure proposed by Newman and Girvan~\cite{newman2004finding}. It is based on the idea that if community structure exists, the density of within-community edges should be much larger than the expected value of that in a randomly connected network with the same community partition and degree distribution. Modularity is defined as the difference between the density of within-community edges and the expected value in the corresponding random graph:
\begin{equation}
    Q=\frac{1}{2M}\sum_{v_i,v_j\in V}[A_{ij}-\frac{k_i\cdot k_j}{2M}]\cdot \delta(c_i, c_j)
\label{12}
\end{equation}
where $c_i$ is the community that $v_i$ belongs to and the $\delta$ function equals to 1 if $c_i=c_j$ or equals to 0 otherwise. Our FPPM further goes through the $N$ partitions in the dendrogram, calculates the modularities of all partitions and picks out the partition $\mathcal{C}_Q$ with the largest modularity. While $\mathcal{C}_Q$ is already a good partition, we find there exist a few terrible cases when go deep into each community. For example, a few vertices located at the border of a very large community or in the common boundary of some great large communities would not be merged into any communities, which is unreasonable. The central panel in Fig. \ref{fig1} shows the emergence of this situation in the karate club graph~\cite{zachary1977information}, where $v_9,v_{11},v_{28}$ are the terrible communities.

We further develop an additional procedure to fix this problem which is inspired by the Label Propagation Algorithm (LPA). LPA is a simple, fast and usually effective community detection method proposed by Raghavan et al~\cite{raghavan2007near}. In LPA, the community of a vertex is determined by the majority community of its neighbors. Taking the similar strategy with LPA, we merge communities whose sizes are smaller than $\theta_c$ into other communities, where $\theta_c$ is the threshold of a reasonable community. In our algorithm, $\theta_c$ is set to 3. For convenience, the communities that are smaller than $\theta_c$ are called small communities. Further, the relevance degree between a small community $C_s$ and a big community $C_b$ is defined as the sum of similarities between nodes in $C_s$ and adjacent nodes in $C_b$:
\begin{equation}
    R_{C_s}(C_b)=\sum_{(v_i, v_j)\in C_s\times C_b\cap E} s_{ij}
\label{13}
\end{equation}
We then merge each small community into a big community with the largest relevance degree.

Our method traverses each communities in $\mathcal{C}_Q$ and picks out all the small communities. The following operations will be repeated until there exists no small community:
\begin{itemize}
    \item Go through small communities and check if the small community is adjacent to big communities.
    \item If there is no adjacent big community, just skip the small community in this round.
    \item Otherwise, calculate the relevance degrees between itself and its adjacent big communities. Then merge it into the big community with the largest relevance degree.
\end{itemize}

The above-mentioned processing is bound to end because of the assumption that the network is connected. When the post-processing finish, we get the final partition.

\subsection{Time complexity}
In the random walk part, we need to calculate $F^{(n)}, n\leq d$ where $d$ is the diameter of the network. As $T$ only has $2M$ non-zero elements, the calculations can be done in $O(dMN)$ by Eq. (\ref{5}). In addition, the similarities that are defined as the weighted sum of the correlations between first passage probabilities can be calculated in $O(dN^2)$. For HAC part, an algorithm called nearest-neighbors chain is implemented, which has time complexity $O(N^2)$\cite{mullner2011modern}. In the post-processing procedure, $N$ values of modularity are needed to be calculated and its time complexity is $O(MN)$. To eliminate the small communities, edges that connect small communities and big communities are no more than $M$ and vertices that need to change communities are no more than $N$, so the time complexity is less than $O(MN)$. Therefore FPPM has the time complexity $O(dN(M+N))$. When the network is sparse, the time complexity is $O(dN^2)$. Further, most real-world networks have very small diameters compared to $N$, the time complexity of FPPM is approximately $O(N^2)$.

\section{Simulations}

\begin{figure*}
    \centering\includegraphics[width=12cm]{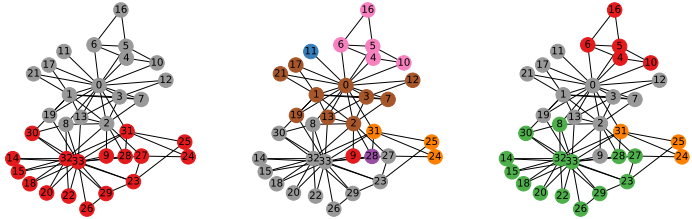}
    \caption{Comparisons between the ground-truth and the communities predicted by FPPM on Karate club graph. The colors of vertices indicate different communities. The left panel shows that there are two communities in the ground-truth. The center panel is the partition $\mathcal{C}_Q$ selected by the maximal modularity, where $v_9, v_{11}, v_{28}$ are terrible communities. In the right panel, FPPM further divides each community into two smaller communities, within which only vertices $v_8$ and $v_9$ are misclassified.}
\label{fig1}
\end{figure*}

In this section, we explore the performance of FPPM on various synthetic and real-world networks. Firstly, we begin with a simple yet typical instance -- the karate club graph (Fig. \ref{fig1}, left panel), which is a friendship network composed of 34 members and possesses clear community structure~\cite{zachary1977information}. FPPM divides each real community into two smaller ones (Fig. \ref{fig1}, right panel), which reveals the existence of hierarchical community structure. In this partition, only agents $v_8$ and $v_9$ are misclassified, and $v_8$ has two friends in the real community yet three friends in the "misclassified" community, which indicates that FPPM actually performs well.

We further examine how FPPM behaves on large-scale networks, including two synthetic and four real-world networks, and compare it with five well-known algorithms including Fastgreedy~\cite{clauset2004finding}, Infomap~\cite{raghavan2007near}, LPA~\cite{raghavan2007near}, Louvein~\cite{blondel2008fast} and Walktrap~\cite{pons2005computing}. Complying with previous studies, we utilize normalized mutual information (NMI) as the quality function to measure the similarity between the ground-truth and the partition delivered by algorithms~\cite{danon2005comparing}.

All the datasets in this paper and the Python code for FPPM can be obtained at \textit{https://github.com/peterwu4084/FPPM}.

\subsection{Planted l-partition benchmark}

\begin{figure}
    \centering\includegraphics[width=12cm]{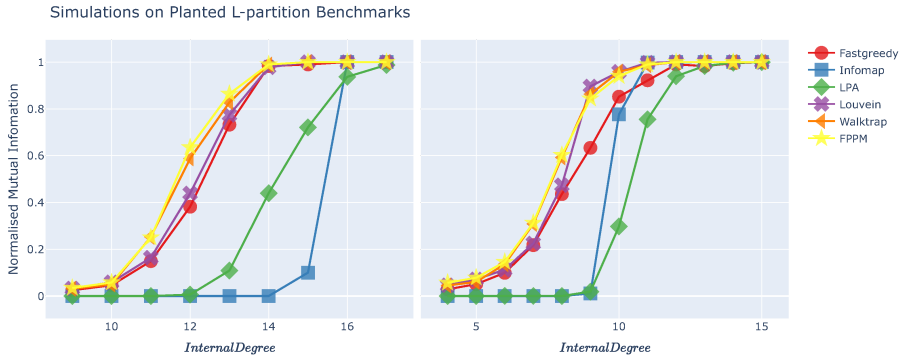}
    \caption{Comparisons on l-partition benchmarks measured by NMI between the ground-truth and the predicted communities as a function of the internal degree $d_{in}$. Parameters: (left panel) $N=64, l=2, \langle k\rangle=18, d_{in} \in \{9,10,...,17\}$. (right panel) $N=128, l=4, \langle k\rangle=16, d_{in} \in \{4, 5, ...,15\}$. Each point is the average of $1000$ simulations.}
\label{fig2}
\end{figure}

The first synthetic datasets are generated by the planted l-partition model~\cite{condon2001algorithms}. This model generates networks with $N=l\cdot N_c$ vertices and $l$ communities with identical size. Vertices are connected with a probability $p_{in}$ if they belong to the same community and $p_{out}$ otherwise. The expected degree of each vertex satisfies $\langle k\rangle=p_{in}\cdot(N_c - 1) + p_{out}\cdot N_c \cdot (l-1)$, where $d_{in}=p_{in}\cdot(N_c-1)$ and $d_{out}=p_{out}\cdot N_c \cdot (l-1)$ are known as the internal degree and the external degree. Here we select two groups of parameters to generate networks. The first group is $N=64, l=2, \langle k\rangle=18, d_{in} \in\{9, 10,...,17\}$ and the second one is $N=128, l=4, \langle k\rangle=16, d_{in}\in\{4,5,...,15\}$. The range of $d_{in}$ is calculate by $p_{in} > p_{out} > 0 $, under which condition it is considered that there exists community structures in the network. For each combination of parameters, we generate $10$ networks and run $100$ independent simulations on each network for all methods.

Fig. \ref{fig2} shows that FPPM outperforms most of the competitors. In the first case (Fig. \ref{fig2} left panel), FPPM and Walktrap work best and FPPM is slightly better than Walktrap. In the second case (Fig. \ref{fig2} right panel), FPPM is as effective as Walktrap and performs best at most values of $d_{in}$, except for $d_{in}=9$ and $10$.

\subsection{LFR benchmark}
\begin{figure}
    \centering\includegraphics[width=12cm]{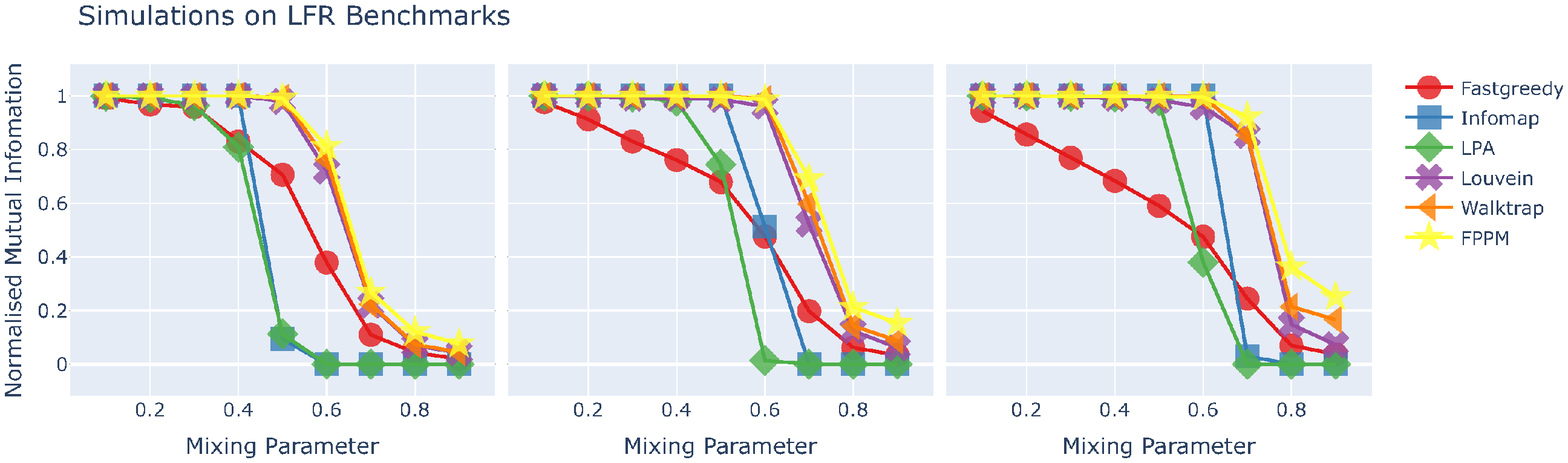}
    \caption{Comparisons on LFR benchmarks measured by NMI between the ground-truth and the predicted communities as a function of the mixing parameter $\mu$. Parameters: $\tau_1=2, \tau_2=1, \langle k\rangle=25, \mu \in \{0.1, 0.2, ..., 0.9\}$. From left to right, $N=250, 500, 1000$ respectively. Each point is the average of 1000 simulations.}
\label{fig3}
\end{figure}

The second synthetic dataset is LFR benchmark~\cite{lancichinetti2008benchmark}, which a better proxy of real-world networks compared with the planted l-partiton benchmark. The model is able to generate networks that have power law degree distribution and community size with exponents $\tau_1$ and $\tau_2$ respectively. The mixing parameter $\mu$ controls the fraction of neighbors from different communities for each vertex. We test our method on LFR networks with $N\in\{250, 500, 1000\}, \mu\in\{0.1,0.2,...,0.9\}, \tau_1=2, \tau_2=1, \langle k\rangle=25$. Similar to the previous settings, ten networks are generated with each combination of parameters and each method is simulated 100 times on each network. Fig. \ref{fig3} shows that the performance of all algorithms drops as $\mu$ increases and FPPM outperforms all the competitors, which exhibits the great power of our method.

\subsection{Real-world datasets}

Finally we apply FPPM on several real-world networks, the details of which are as follows :
\begin{itemize}
    \item Polblogs~\cite{adamic2005political}: a network of hyperlinks between blogs about US politics. Each vertex is labeled by an integer to indicate liberal (0) or conservative (1).
    \item Polbooks~\cite{polbooks}: a network of books about US politics published around 2004 and sold on Amazon. Edges between books mean frequent co-purchasing by the same customers. Each vertex is labeled by "l" (liberal), "n" (neutral) or "c" (conservative).
    \item Cora~\cite{rossi2015network}: a directed network consists of 2708 scientific publications classified into one of seven classes. An edge from $v_i$ to $v_j$ indicates that $v_i$ cites $v_j$.
    \item Citeseer~\cite{rossi2015network}: a directed network extracted from the Citeseer digital library which contains 3264 vertices, representing publications from six domains. Similar to Cora, edges represent citation relationships.
\end{itemize}

For each network, only the giant component is kept and the direction of edges is ignored. The statistical properties of these real-world networks and the performances of different algorithms are shown in Table. \ref{table1}. Each performance value is the average of $100$ independent simulations. Results show that FPPM performs best in all situations, which proves that FPPM can provide valuable predictions of community structure on real-world networks.

\begin{table}[]
\Large
    \centering
    \resizebox{\textwidth}{!}{%
    \begin{tabular}{ccccclcccccc}
    \hline
    \multirow{2}{*}{Network} & \multirow{2}{*}{N} & \multirow{2}{*}{M} & \multirow{2}{*}{$\langle k\rangle$} & \multirow{2}{*}{Communities} &  & \multicolumn{6}{c}{Normalised Mutual Information}                          \\ \cline{7-12}
                             &                    &                    &                    &                              &  & Fastgreedy & Infomap  & LPA      & Louvein  & Walktrap & FPPCD             \\ \hline
    Polblogs                 & 1222               & 16717              & 27.36              & 2                            &  & 0.654091   & 0.485194 & 0.680634 & 0.644623 & 0.646807 & \textbf{0.694281} \\
    Polbooks                 & 105                & 441                & 8.40               & 3                            &  & 0.530814   & 0.493454 & 0.556470 & 0.512133 & 0.542748 & \textbf{0.564378} \\
    Cora                     & 2485               & 5069               & 4.08               & 7                            &  & 0.460671   & 0.416478 & 0.431760 & 0.474595 & 0.442702 & \textbf{0.495471} \\
    Citeseer                 & 2110               & 3668               & 3.48               & 6                            &  & 0.343909   & 0.341153 & 0.339459 & 0.329998 & 0.354050 & \textbf{0.372667} \\ \hline
    \end{tabular}%
    }
    \caption{The statistical properties of the giant components of real-world networks and the performances of different algorithms measured by NMI between the ground-truth and the predicted communities.}
    \label{table1}
    \end{table}

\section{Conclusion}
Community detection aims to classify similar vertices into the same class on complex networks, which not only is helpful for understanding the structural characteristics, but also can be applied into many dynamical processes, such as mitigating the intensity of epidemics, controlling rumors and identifying the core area of brain networks. Great efforts have been made in developing algorithms for detecting communities. In particular, random walk has been proven to be an efficient and classic method of mining community structures and is widely used, which leads to the explosion of many well-known algorithms. Nevertheless, most of the existing methods have one of the two major defects: (1) using the average steps of all possible random walks, which is often too large to distinguish the structural information; (2) using a pre-assigned length of steps, which may ignore useful community information at other step lengths.

In this paper, we propose a novel community detection algorithm (FPPM) based on the first passage probabilities. The diameter of the network is used as the maximal step length of random walks which can be adaptively appropriate to different networks. We then introduce a new similarity measure via the sum of weighted correlation between first passage probabilities at multiple times, which considers the complete structural information rather than just one fixed-length step. Furthermore, the vertices are grouped into communities by using the hierarchical clustering, and the best division can be achieved by choosing the maximal modularity value. Finally, similar to LPA, a post-processing procedure is conducted to remove the unreasonable small community which further improves the accuracy of community detection results. Surprisingly, numerous simulations show that FPPM performs better than several classic methods on both synthetic benchmarks and real-world networks, which indicates the great power and the universality of our algorithm.

\section*{Acknowledgement}
This work is supported by Program of National Natural Science Foundation of China Grant No. 11871004, 11922102.

\section*{References}


\end{document}